\documentclass[12pt,epsf]{article}

\begin{document}

\begin{titlepage}

\begin{flushright}
IUHET-460\\
hep-th/0311200
\end{flushright}
\vskip 2.5cm

\begin{center}
{\Large \bf Failure of Gauge Invariance in the Nonperturbative Formulation of
Massless Lorentz-Violating QED}
\end{center}

\vspace{1ex}

\begin{center}
{\large B. Altschul\footnote{{\tt baltschu@indiana.edu}}}

\vspace{5mm}
{\sl Department of Physics} \\
{\sl Indiana University} \\
{\sl Bloomington, IN 47405 USA} \\

\end{center}

\vspace{2.5ex}

\medskip

\centerline {\bf Abstract}

\bigskip

We consider a Lorentz-violating modification to the fermionic Lagrangian of QED
that
is known to produce a finite Chern-Simons term at leading order. We compute the
second order correction to the one-loop photon self-energy in the massless case
using an exact propagator and a nonperturbative formulation of the theory.
This nonperturbative theory assigns a definite value to the coefficient
of the induced Chern-Simons term; however, we find that the theory fails to
preserve gauge invariance at higher orders. We conclude that the specific
nonperturbative value of the Chern-Simons coefficient has no special
significance.

\bigskip

\end{titlepage}

\newpage

There has been much recent interest in the possibility of adding CPT- and
Lorentz-violating terms to the Lagrangians of quantum field
theories~\cite{ref-kost1,ref-kost2,ref-kost3,ref-kost4}. These terms may
arise
from violations of these symmetries at the Planck scale. There are many strong
experimental constraints on Lorentz-violating corrections to the standard
model, and such corrections must generally be small. However, the subject of
Lorentz violation in quantum field theory is still of great theoretical and
experimental interest.

The simplest perturbatively nontrivial correction to the fermion sector of
quantum electrodynamics
involves the addition of a CPT-violating axial vector term to the action. There
has been a great deal of interest in the effect of such a term on the radiative
corrections~\cite{ref-kost4,ref-jackiw1,ref-victoria1,ref-chung1,ref-coleman,
ref-chung2,ref-victoria2,ref-kost5}
to the theory. At one
loop order, the theory generates a term of the Chern-Simons
form ${\cal L}_{CS}=\frac{1}{2}(k_{CS})_{\mu}\epsilon^{\mu\alpha\beta\gamma}
F_{\alpha\beta}A_{\gamma}$~\cite{ref-jackiw2,ref-schonfeld,ref-carroll1}.
Astrophysical measurements constrain the physical coefficient $(k_{CS})_{\mu}$
to be very small~\cite{ref-carroll1,ref-carroll2,ref-goldhaber}.
However, while the radiatively induced value of $(k_{CS})_{\mu}$ is
unambiguously finite, it is also completely undetermined (and possibly
vanishing); its value depends upon how the theory is regulated.
Moreover, since an arbitrary tree-level contribution may also be added to the
radiatively induced term, the calculated value of the induced Chern-Simons
coefficient can have no experimental significance.

There is, however, one particular value of the coefficient that appears to
enjoy a special status. It is pointed out in~\cite{ref-jackiw1} that if the
theory is defined nonperturbatively in the CPT-violating axial vector
interaction, then there is a precise value for the induced coefficient. This
raises the question of whether this nonperturbative formulation has any
special significance. In this paper, we answer that question, at least for the
case of massless fermions, by analyzing higher-order contributions to the
photon self-energy. We find that the nonperturbative formalism cannot be
consistently applied to the calculation of the radiatively induced, CPT-even
corrections to the theory. A nonperturbative regularization of the sort used
in~\cite{ref-jackiw1} leads unavoidably
to a violation of the Ward identity that enforces the transversality of the
vacuum
polarization. Alternatively, while the use of dimensional regularization leads
to
unambiguously transverse CPT-even terms, this regularization restores the
complete ambiguity in the CPT-odd terms, because $\gamma_{5}$ does not have a
unique dimensional extension.

We shall introduce the massless theory and exhibit an extremely simple
rationalization of the exact propagator. We then move on to
calculate the one-loop photon self-energy $\Pi^{\mu\nu}(p)$. We shall interpret
our results with the help of
an analogy to a simpler theory, in which apparent violations of the Ward
identity $p_{\mu}\Pi^{\mu\nu}(p)=0$ can be eliminated by a change in the
regularization; however, the corresponding
change in regularization for the theory of interest is fundamentally
nonperturbative. We conclude with a discussion of the implications of
our result.

The Lagrange density for our theory (including a possible mass term) is
\begin{equation}
\label{eq-L}
{\cal L}=-\frac{1}{4}F^{\mu\nu}F_{\mu\nu}+\bar{\psi}(i\!\!\not\!\partial-m-
e\!\!\not\!\!A\,-\!\not\!b\gamma_{5})\psi.
\end{equation}
However,
we shall be concerned here only with the $m=0$ case. Although we can eliminate
$b$ from the massless Lagrangian by making the chiral transformation
\begin{equation}
\label{eq-psiredef}
\psi\rightarrow e^{-i\gamma_{5}b\cdot x}\psi,\,\bar{\psi}\rightarrow\bar{\psi}
e^{-i\gamma_{5}b\cdot x},
\end{equation}
this transformation is
anomalous and does not leave the gauge invariantly regulated fermionic measure
invariant~\cite{ref-fujikawa}. However, we should keep in mind that, if we
account correctly for the anomaly associated with (\ref{eq-psiredef}), we can
eliminate $b$ from the rest of the theory.

There are several reasons for considering only the massless case. Setting $m=0$
simplifies the algebra in the calculation of the self-energy, but this is
a minor point. There are two other, more important reasons. The first is the
chiral symmetry mentioned above. The existence of this symmetry significantly
simplifies our discussion; in particular, it allows us to construct a clear
analogy that will illuminate the origin of the difficulties we encounter. The
second reason is more subtle. We shall use power-counting arguments to
determine the structure of the $b$-dependence of the vacuum polarization. In
order to apply these arguments, we must suppose that the theory can be expanded
in a power series in $b$. If there exists a nonvanishing momentum scale $m^{2}$
in the theory, it is conceivable that the power series description might break
down at the scale $b^{2}\sim m^{2}$. This actually occurs in the calculation
of the Chern-Simons term~\cite{ref-victoria1,ref-chung1}. To avoid similar
problems here, we set $m=0$.

The exact fermion propagator for the massive theory is
\begin{equation}
S(k)=\frac{i}{\!\not\!k-m\,-\!\not\!b\gamma_{5}}.
\end{equation}
When $m=0$, this
is most easily rationalized by breaking it into two terms, corresponding
to the two eigenvalues of $\gamma_{5}$. Doing this, we have
\begin{eqnarray}
S(k) & = & \frac{i}{\!\not\!k\,-\!\not\!b}\frac{1-\gamma_{5}}{2}+\frac{i}{\!
\not\!k\,+\!\not\!b}\frac{1+\gamma_{5}}{2} \nonumber\\
\label{eq-propagator}
& = & i\left[\frac{\!\not\!k\,-\!\not\!b}{(k-b)
^{2}}\frac{1-\gamma_{5}}{2}+\frac{\!\not\!k\,+\!\not\!b}{(k+b)^{2}}
\frac{1+\gamma_{5}}{2}\right].
\end{eqnarray}
This rationalization of the propagator is substantially simpler than other
versions.
The numerators have fewer Dirac matrices, and the denominators are simpler as
well. The presence
of the right- and left-handed projectors will also simplify the algebra.

In particular, the one-loop self-energy,
\begin{equation}
\Pi^{\mu\nu}(p)=-ie^{2}{\rm tr}\int\frac{d^{4}k}{(2\pi)^{4}}\gamma^{\mu}S(k)
\gamma^{\nu}S(k+p),
\end{equation}
may be simplified in the massless case to
\begin{eqnarray}
\Pi^{\mu\nu}(p) & = & \frac{ie^{2}}{2}{\rm tr}\int\frac{d^{4}k}{(2\pi)^{4}}
\left[\frac{\gamma^{\mu}(\!\not\!k\,+\!\not\!b\,)\gamma^{\nu}(\!\not\!k\,+\!
\not\!p\,+\!\not\!b\,)+
\gamma^{\mu}(\!\not\!k\,+\!\not\!b\,)\gamma^{\nu}(\!\not\!k\,+\!
\not\!p\,+\!\not\!b\,)\gamma_{5}}
{(k+b)^{2}(k+p+b)^{2}}\right.\nonumber\\
\label{eq-selfE}
& & +\left.\frac{\gamma^{\mu}(\!\not\!k\,-\!\not\!b\,)\gamma^{\nu}(\!\not\!k\,
+\!\not\!p\,-\!\not\!b\,)-
\gamma^{\mu}(\!\not\!k\,-\!\not\!b\,)\gamma^{\nu}(\!\not\!k\,+\!
\not\!p\,-\!\not\!b\,)\gamma_{5}}
{(k-b)^{2}(k+p-b)^{2}}\right].
\end{eqnarray}
Through an analysis of the structure of (\ref{eq-selfE}), we may learn a great
deal about the nature of the nonperturbative theory.

We observe that all the information necessary for the calculation of $\Pi^{\mu
\nu}(p)$ is contained in the function
\begin{equation}
\label{eq-fdef}
f_{\alpha\beta}(p,b)=\int\frac{d^{4}k}{(2\pi)^{4}}\frac{(k-b)_{\alpha}
(k+p-b)_{\beta}}{(k-b)^{2}(k+p-b)^{2}}.
\end{equation}
The self-energy involves symmetric and antisymmetric sums $f_{\alpha\beta}
(p,b)\pm f_{\alpha\beta}(p,-b)$ contracted with tensors in $(\mu,\nu,
\alpha,\beta)$.
The terms of $f_{\alpha\beta}(p,b)$ that are even in $b$ give rise to
contributions to $\Pi^{\mu\nu}(p)$ with different Lorentz structure than the
terms
that are odd in $b$, since the odd terms involve a trace over $\gamma_{5}$.
We might conclude that, because of the differences in their Lorentz structures,
the two types of terms will need to be regulated differently.
However, as a formal object, $f_{\alpha\beta}(p,b)$ still contains
everything needed to determine the one-loop self-energy.

In fact, the nonperturbative viewpoint requires that the same regulator
be used for all the terms, regardless of whether they are even or odd in $b$.
Since there is
only a single Feynman diagram in the nonperturbative formulation, a truly
nonperturbative calculation would involve a single evaluation of
$f_{\alpha\beta}(p,b)$ to all orders in $b$ using a unique regularization
prescription. It would not be consistent to use the nonperturbative
regularization for the $b$-odd terms and a different regulator for the even
terms. (It is possible that in a more fundamental theory, we may be
required to use
a regulator that does treat the even and odd terms differently; however,
this is just speculation.)
We shall therefore use the same methods used in~\cite{ref-jackiw1} to
fix the coefficient of the induced Chern-Simons term to determine the
higher-order, CPT-even contributions to the self-energy.

If we shift the integration variable $k\rightarrow k-b$ in $f_{\alpha\beta}
(p,b)$, then the integrand becomes $b$-independent. Since the integral is
superficially quadratically
divergent, the surface term generated by the shift is at most quadratic in $b$.
Therefore, there are no contributions to the self-energy that are higher than
second order in $b$. This result has been previously demonstrated
for the $b$-odd terms, and it might be expected on
dimensional grounds for the $b$-even terms as well, since there is no mass
scale in the problem. However, the ease with which it has been demonstrated
here shows the usefulness of the propagator (\ref{eq-propagator}).

The ${\cal O}(b^{0})$ part of $f_{\alpha\beta}(p,b)$ gives the usual
QED photon self-energy. The ${\cal O}(b)$ contribution to $\Pi^{\mu\nu}(p)$ has
also been calculated; this is the Chern-Simons term. In a specific
nonperturbative formalism, which is presented
in~\cite{ref-jackiw1,ref-victoria1,ref-chung1},
this term has a fixed value, but more generally---in particular,
if the theory is defined perturbatively in $b$---its value is undetermined and
regularization-dependent~\cite{ref-victoria2}.
However, this ambiguity and the structure of this term in general are well
understood.
We shall therefore focus our attention on the ${\cal O}(b^{2})$ terms.

If we were interested only in finding the ultraviolet divergent part of
$f_{\alpha
\beta}(p,b)$ at second order in $b$, we could simply shift the integration as
outlined above and evaluate the resulting integral; since
the initial integral is superficially quadratically divergent, the
${\cal O}(b^{2})$ part of the surface term accompanying the shift is
ultraviolet finite. We would find that the ultraviolet divergent
part of $f_{\alpha\beta}(p,b)$ vanishes at ${\cal O}(b^{2})$.
However, if we wish to calculate the ultraviolet finite contributions (which
are derived entirely from the surface term accompanying the integration shift)
as well, we must take more care.

We shall evaluate the ${\cal O}(b^{2})$ terms in $f_{\alpha\beta}(p,b)$
by a direct expansion of the integrand in powers of $b$.
The integral that appears at second order in $b$ is finite when the
integration is performed symmetrically. This is the correct prescription for
performing the integration in the nonperturbative formalism, because the same
property
(observer Lorentz invariance) is being used to fix the value of the vacuum
polarization at both ${\cal O}(b)$ and ${\cal O}(b^{2})$. [Of course, the same
technique cannot be used to deal with the ${\cal O}(b^{0})$ term in the
photon self-energy if gauge invariance is to be preserved. We shall set this
formal difficulty aside, however, since the ${\cal O}(b^{0})$ is necessarily
divergent in any regularization scheme and is thus qualitatively different from
the higher-order terms.]

We begin our calculation by writing
\begin{equation}
f_{\alpha\beta}(p,b)=\int\frac{d^{4}k}{(2\pi)^{4}}
\exp\left(-b_{\gamma}\frac{\partial}{\partial k_{\gamma}}\right)\frac
{k_{\alpha}}{k^{2}}\frac{k'_{\beta}}{(k')^{2}},
\end{equation}
where we have defined $k'=k+p$. The portion of this expression that is
quadratic in $b$, which we shall denote as $h_{\alpha\beta}(p,b)$, is
\begin{eqnarray}
h_{\alpha\beta}(p,b) & = &  \frac{1}{2}b_{\gamma}b_{\delta}
\int\frac{d^{4}k}{(2\pi)^{4}}\frac{\partial}{\partial k_{\gamma}}\frac
{\partial}{\partial
k_{\delta}}\frac{k_{\alpha}}{k^{2}}\frac{k'_{\beta}}{(k')^{2}} \\
\label{eq-h}
& = & \frac{1}{96\pi^{2}}(b_{\alpha}b_{\beta}-g_{\alpha\beta}b^{2}).
\end{eqnarray}

Transversality of the
vacuum polarization requires that $(2p^{\alpha}g^{\nu\beta}-p^{\nu}g^{\alpha
\beta})h_{\alpha\beta}(p,b)=0$. This condition does not hold; even though
$h_{\alpha\beta}(p,b)$ is unambiguously finite, it still violates the Ward
identity. Gauge invariance is broken, so the theory
becomes nonrenormalizable, and we encounter divergences, which render the
theory undefined.

To understand how this violation of gauge invariance arises, it is useful to
consider an analogy. For a theory with Lagrange density
\begin{equation}
{\cal L}'=-\frac{1}{4}F^{\mu\nu}F_{\mu\nu}+\bar{\psi}(i\!\!\not\!\partial-
e\!\!\not\!\!A\,-\!\not\!a\,)\psi,
\end{equation}
there is a field rescaling similar to (\ref{eq-psiredef}),
\begin{equation}
\label{eq-aredef}
\psi\rightarrow e^{-ia\cdot x}\psi,\,\bar{\psi}\rightarrow e^{ia\cdot x}
\bar{\psi},
\end{equation}
which eliminates $a$ from the theory. Moreover, this rescaling is not
anomalous. However, we may choose not to eliminate $a$ from the
Lagrangian. If we consider the theory directly as defined by ${\cal L}'$ and
attempt to calculate the photon self-energy, we will immediately be led to
an evaluation of $f_{\alpha\beta}(p,a)$. The violation of the Ward
identity in this instance is clearly an artifact of our unconventional choice
of momentum coordinates. A shift in the integration variable $k\rightarrow
k-a$ eliminates the problematic term. This shift is
precisely equivalent to the field redefinition (\ref{eq-aredef}).

Returning to the axial vector theory (\ref{eq-L}), it seems now that the
correct solution to our difficulties would be to shift the origin of the
integration in (\ref{eq-fdef}) so that $h_{\alpha\beta}(p,b)$ is set to zero.
While we believe that this is the physically correct way of regulating this
theory, it is inconsistent with the nonperturbative formalism. In the
vector theory with $a$, there arises in the calculation of $\Pi^{\mu\nu}(p)$
only a single term of the form $f_{\alpha\beta}(p,a)$. However, in the axial
vector theory, we encounter the sum $f_{\alpha\beta}(p,b)+f_{\alpha\beta}(p,
-b)$. The essence of the nonperturbative formulation is that both terms in
(\ref{eq-selfE}) must be regulated in the same way. We are allowed only a
single shift in the integration variable $k\rightarrow k+q$. This shift
transforms $f_{\alpha\beta}(p,b)+f_{\alpha\beta}(p,-b)\rightarrow
f_{\alpha\beta}(p,q+b)+f_{\alpha\beta}(p,q-b)\neq0$.
We see that in order to eliminate
the Ward-identity-violating surface term, we must be free to shift the
integrations in different terms by different amounts, which is equivalent to
defining the theory perturbatively. [Although the
theory is massless, a chiral shift $k\rightarrow k+\gamma_{5}q$ cannot be
implemented in a fashion that allows us to retain a unique result at ${\cal O}
(b)$, for reasons we shall outline below.]
In the perturbative formulation, we treat
the term $-\bar{\psi}\!\not\!b\gamma_{5}\psi$ in ${\cal L}$ as defining a new
vertex of the theory. We are then free to shift the integrations independently
in the evaluations of different Feynman diagrams, and this allows us to enforce
gauge invariance.

Although we are working with a massless theory,
we are not allowed to make naive chiral shifts in the integration variable $k$
if we are to retain the special nonperturbative value for $(k_{CS})_{\mu}$,
because the corresponding transformation (\ref{eq-psiredef}) is anomalous. To
properly account for the anomaly, we must use the functional integral
formalism. The Fujikawa determinant~\cite{ref-fujikawa} accompanying the shift
then reproduces the correct nonperturbative value of the induced Chern-Simons
coefficient; however, there is another, completely undetermined contribution
that arises from the ambiguity in the definition of the axial current
operator~\cite{ref-chung3}. So it is impossible to shift the integrations
separately for the left- and right-handed components of $\psi$ without giving
up the uniqueness of the ${\cal O}(b)$ result.

It is the necessity of using a single regulator for both terms in
(\ref{eq-selfE}) that gives rise to a unique specification of the induced
Chern-Simons term at ${\cal O}(b)$. However, it proved impossible to regulate
the ${\cal O}(b^{2})$ terms in the same fashion without
violating the Ward identity.
Therefore, there seems to be no special significance to the nonperturbative
value
of the induced Chern-Simons term. This is in some ways unsurprising.
Although there are stability problems for
theories with nonvanishing Chern-Simons coefficients,
we are always free to add an additional Chern-Simons term to the bare Lagrangian
so as to make the total coefficient zero.

The existence of a surface term that violates gauge invariance at
${\cal O}(b^{2})$ is perhaps also unsurprising, given what is known about the
behavior of the nonperturbative
theory at ${\cal O}(b)$. In the massless case, the only contribution to the
induced Chern-Simons coefficient comes from the surface term. The associated
induced Lagrange density is not gauge invariant; however, because the density
necessarily involves $\epsilon^{\mu\nu\alpha\beta}$, the Ward identity is
preserved and the integrated action remains gauge invariant. There is no such
restriction on the form of the induced term at ${\cal O}(b^{2})$, and without
the protection of a specific structure for the self-energy, the
gauge invariance of the action is lost. At each order, the surface term simply
violates gauge invariance in the strongest way allowed by its tensor structure.

Finally, we must discuss the possibility of dimensional regularization. A
dimensional regulator preserves gauge invariance at all orders in $b$ and sets
$h_{\alpha\beta}(p,b)=0$. Moreover, it solves the formal problems associated
with the nonperturbative evaluation of the ${\cal O}(b^{0})$ contributions,
since it allows us to
regulate all the terms, even the divergent ones, in the same fashion. However,
dimensional regularization also restores the complete ambiguity of the induced
Chern-Simons term. (This is closely analogous to the situation we encountered
when using the functional integral formalism and a Fujikawa regulator.)
The $b$-odd terms involve $\gamma_{5}$, which does not have
a unique extension to $4-\epsilon$ dimensions. Any extension that commutes with
$4-n\epsilon$ $\gamma$-matrices (for arbitrary $n$) will have the correct
limit as $\epsilon\rightarrow0$, and each extension will give a different result
for the Chern-Simons coefficient. Moreover, it is not possible to determine the
correct extension from other sectors of the theory. The dimensional extension
of the $\gamma_{5}$ appearing
in $-\bar{\psi}\!\not\!b\gamma_{5}\psi$ need not be the same
as the dimensional extensions of the $\gamma_{5}$ appearing in the chiral gauge
sector, for example; these are entirely distinct operators, which may behave
differently under dimensional regularization. Therefore, while a dimensional
regularization prescription may be used to implement a completely
nonperturbative formulation of the theory, it also renders the coefficient of
the induced Chern-Simons term completely undetermined.

In this paper, we have demonstrated that the ambiguity in the value of the
induced Chern-Simons term for the theory defined by (\ref{eq-L}) is
unavoidable.
The specific value predicted by the nonperturbative formulation
can have no special significance, because the nonperturbative regularization of
the theory used in~\cite{ref-jackiw1} cannot be extended to higher orders in $b$
without violating the
Ward identity; and therefore, even if it contains no tree-level contribution,
the coefficient of Chern-Simons term in the effective action can be fixed only
by experiment. Although this result was derived only in the limit of massless
fermions, the massive theory may well behave similarly.

\section*{Acknowledgments}
The author is grateful to R. Jackiw and V. A. Kosteleck\'{y} for many helpful
discussions.
This work is supported in part by funds provided by the U. S.
Department of Energy (D.O.E.) under cooperative research agreement
DE-FG02-91ER40661.

\end{document}